\documentclass{iopjournal}
\usepackage{upgreek}
\usepackage{amsmath}
\AtEndEnvironment{thebibliography}{
\bibitem{data}
        B.~Brereton, S.~Hait, C.~J.~Kinane, F.~Maccherozzi, A.~Yagmur, T.~A.~Moore, S.~Sasaki, S.~Langridge,
and C.~H.~Marrows, Research Data Leeds, DOI:tbc, 2026

\bibitem{supp}
    See Supplementary Material at [URL will be inserted by publisher] which includes references [48 - 51]

\bibitem{Stull1947} Stull D R 1947 \textit{Industrial \& Engineering Chemistry} \textbf{39} 517–550

\bibitem{Cafaro1984} Cafaro M L, Bardi G and Piacente V 1984 \textit{Journal of Chemical \& Engineering Data} \textbf{29} 78–80

\bibitem{Momma2011} Momma K and Izumi F 2011 \textit{J. Appl. Crystallogr.} \textbf{44} 1272–1276

\bibitem{Xue2013}  Xue L, Zhou P, Zhang C X, He C Y, Hao G L, Sun L Z and Zhong J X 2013 \textit{AIP Advances} \textbf{3}
052105
}

\begin{document}

\newcommand{\BS}{Bi$_2$Se$_3$}
\newcommand{\BT}{Bi$_2$Te$_3$}
\newcommand{\BX}{Bi$_2$X$_3$}
\newcommand{\AM}{A/m$^{2}$}
\newcommand{\degc}{\,$^{\circ}$C}
\newcommand{\um}{$\upmu$m}

\articletype{Paper} %	 e.g. Paper, Letter, Topical Review...

%\title{Improved Domain Structure in Annealed Topological Insulator - Magnetic Multilayer Heterostructures}
%\title{Annealing Domain-Hosting Topological Insulator - Magnetic Multilayer Heterostructures}
\title{Effect of Annealing on Maze Domains in Topological Insulator - Magnetic Multilayer Heterostructures}

\author{Benjamin~A.~Brereton$^{1,2,3}$\orcid{0009-0001-2606-3144}, Soumyarup~Hait$^1$\orcid{0000-0002-1861-3353}, Christy~J.~Kinane$^2$\orcid{0000-0002-1185-0719}, Francesco~Maccherozzi$^3$\orcid{0000-0003-4074-2319}, Ahmet~Yagmur$^1$\orcid{0000-0002-6042-7730}, Satoshi~Sasaki$^1$\orcid{0000-0002-8915-6735}, Thomas~A.~Moore$^1$\orcid{0000-0001-6443-2556}, Sean~Langridge$^2$\orcid{0000-0003-1104-0772}, and Christopher~H.~Marrows$^{1 \,*}$\orcid{0000-0003-4812-6393}}

\affil{$^1$ School of Physics and Astronomy, University of Leeds, Leeds LS2 9JT, United Kingdom}

\affil{$^2$ ISIS Neutron and Muon Source, STFC Rutherford Appleton Laboratory, Harwell Science and Innovation Campus,
Chilton, Oxfordshire OX11 0QX, United Kingdom}

\affil{$^3$ Diamond Light Source, Harwell Science and Innovation Campus, Chilton, Oxfordshire OX11 0DE, United Kingdom}

\affil{$^*$Author to whom any correspondence should be addressed.}

\email{c.h.marrows@leeds.ac.uk}

\keywords{Topological Insulators, Magnetic Multilayers, Polarised Neutron Reflectometry, Photoemission Electron Microscopy}

\begin{abstract}

Combining topological insulators with chiral magnetic multilayers offers the opportunity to greatly enhance the spin-orbit torque (SOT) driven manipulation of hosted spin textures, however this also exposes devices to the often poor temperature stability of the commonly used, Bi$_2$X$_3$-based (X is a chalcogenic element), topological insulator films. In this work we demonstrate through \textit{in-situ}-annealing and polarised neutron reflectometry that the multilayer at a \BS\ interface remains magnetically robust up to a temperature of 350\degc: even as the \BS\ epilayer structurally deteriorates. Through x-ray photoemission electron microscopy, an intermediate annealing temperature of $\sim$ 200\degc\ was found to improve the domain ordering of the multilayer, allowing for field-induced skyrmion nucleation, all whilst sustaining minimal damage to the topological insulator.
\end{abstract}

\section{Introduction}

Topological insulators (TIs) are a class of materials that offer to substantially reduce the high current densities often required for spin orbit torque (SOT) driven control of magnetism in next-generation spintronic devices. Unlike spin currents generated through the spin Hall effect in heavy metals, spin polarisation is instead generated by current flowing in fully spin-polarised topological surface states (TSS) produced from an inversion of valence and conduction bands in the material bulk \cite{Xia2009, Li2014, Shiomi2014, Zhang2009}. These surface states are robust against backscattering and yield a measurably higher effective spin Hall angle than classic spin Hall materials such as Pt, Ta or $\beta$-W \cite{Mellnik2014,Fan2016,Dc2018}. Previous studies have demonstrated both how TIs can measurably exert giant damping-like torques on neighbouring ferromagnets as well as drive magnetisation switching \cite{Zhang2025,Fan2014,Wang2017,Han2017,Pan2022}. That TIs have been shown capable of outperforming heavy metals as a SOT source raises the possibility that other spin Hall driven physics should be similarly improved with TI inclusion with most interest being that of spin texture manipulation.

Magnetic skyrmions, and the magnetic multilayers (MMLs) in which they are known to form, have long been of interest for potential low energy, non-volatile data storage \cite{Heinze2011,Moreau,Woo2016,Chen2015,Soumyanarayanan2017,Jiang2015}. Skyrmions can be nucleated and transported through the application of SOTs from the spin Hall effect \cite{Woo2016,Hamamoto2015,Romming2013,Jiang2017,Zeissler2020, Buttner2017}, raising instead the possibility of TSS-driven skyrmion dynamics. % Combining these two material classes hence offers to use their respective strengths to further reduce current density requirements.
Previous work demonstrated how such topological insulator / magnetic multilayer heterostructures can be constructed with minimal structural or magnetic changes to the overlaying multilayer film \cite{Brereton2026}. In that case, the commonly used TI \BS\ was used: a van der Waals material known to vertically stack in quintuple layers which causes distinct terracing at the TI surface. These films were shown to host zero field maze domains suitable for hosting topological spin textures when \BS\ film growth was adjusted to minimise such terracing. Subsequent measurement demonstrates a quantitative increase in the damping-like torque exerted on the MML from insertion of a \BS\ epilayer \cite{Hait2026}. 

%However so far annealing has not been attempted for these heterostructures despite it being a commonly used in the optimisation and modification of thin film properties as well as in previous demonstrations of TI-ferromagnet switching devices.

\BS\ is a material whose above-room temperature properties are already well studied, in large part for its use as a thermoelectric material \cite{Bayaz2003,Mishra1997, Kadel2010}. %Temperature is additionally vital in optimising crystal growth for the topological properties of \BS. Typical ideal growth and / or annealing temperatures under UHV conditions lie in the 200 - 350\degc\ range for MBE  and CVD \cite{ Hellerstedt2014, Brom2016} for optimised epitaxial growth. These ranges roughly match the maximum recorded operating temperature for a \BS-based device . 
Careful consideration and optimisation of film temperature is vital in the thin film growth of \BS\ with a post-growth annealing step often being used \cite{ Bansal2011, He2011}. Increasing temperature, however, particularly post-growth, risks damage to the \BS\ and its electrical transport properties through various mechanisms such as crystal decomposition, surface oxidation and Se loss arising beyond $\sim 200$\degc\ \cite{Puthran2024,Sapkota2021}. This may be problematic for TI-MML heterostructures both as annealing is often used to anneal MMLs post-growth -- often either to increase the effective perpendicular magnetic anisotropy $K_{\text{eff}}$ or the interfacial Dzyaloshinskii-Moriya interaction (iDMI) \cite{Bandiera2012, Rodmacq2009, Cao2020, Kim2019, Gweon2018} -- as well as limiting the temperature range devices can operate due to current-induced Joule heating. TI-metal devices should be expected to be particularly prone to heating due to large parasitic currents that shunt through the low resistance metal films creating increased heat due to the $I^2$ dependence of power dissipation. Knowledge of the heterostructure's response to heating for optimisation and performance is hence critical. 

Here we demonstrate, by establishing an optimal temperature range, how an annealing step can be implemented to improve the ordering of the magnetic domain structure whilst maintaining the structural integrity of the TI epilayer.
In proving the thermal stability of sheet films, we further demonstrate the heterostructures suitability for integration into next-generation spintronic devices.

\section{Method}

In this work we investigate the effect of annealing on a \BS(20)/Ta(1.5)/[Pt(0.8)/CoB(0.6)/Ru(0.5)]$_{\times 6}$/Ru(0.5) stack with perpendicular magnetic anisotropy (PMA), where all given layer thicknesses are in nm. The \BS\ epilayer was grown by molecular beam epitaxy (MBE) on a sapphire substrate before being transferred to an adjoining chamber under UHV to grow the metal films by DC magnetron sputtering. A more detailed description of the sample growth and optimisation can be found in previous work \cite{Brereton2026}. A schematic of the sample stack can be seen in Fig.~\ref{fig:intro}(a). The sample was annealed \textit{in situ} on the POLREF beamline at the ISIS neutron and muon source with polarised neutron reflectivity (PNR) data taken after each anneal cycle. Samples were annealed under a rough vacuum where $P \sim 10^{-3}$~mbar. Here the process pressure remains above the estimated \BS\ vapour pressure up to the furnace temperature limit of 350\degc\ \cite{supp}. The sample was annealed in 50\degc\ increments for 5 minutes at each temperature set-point up to 350\degc\ before returning to room temperature for measurement. Measurements were taken with a 0.7~T applied in-plane field to saturate the sample along the magnetic hard axis.

\section{Results}
\subsection{Structural and Magnetic Effects of Annealing}

\begin{figure}[t]
    \centering
    \includegraphics[width=14cm]{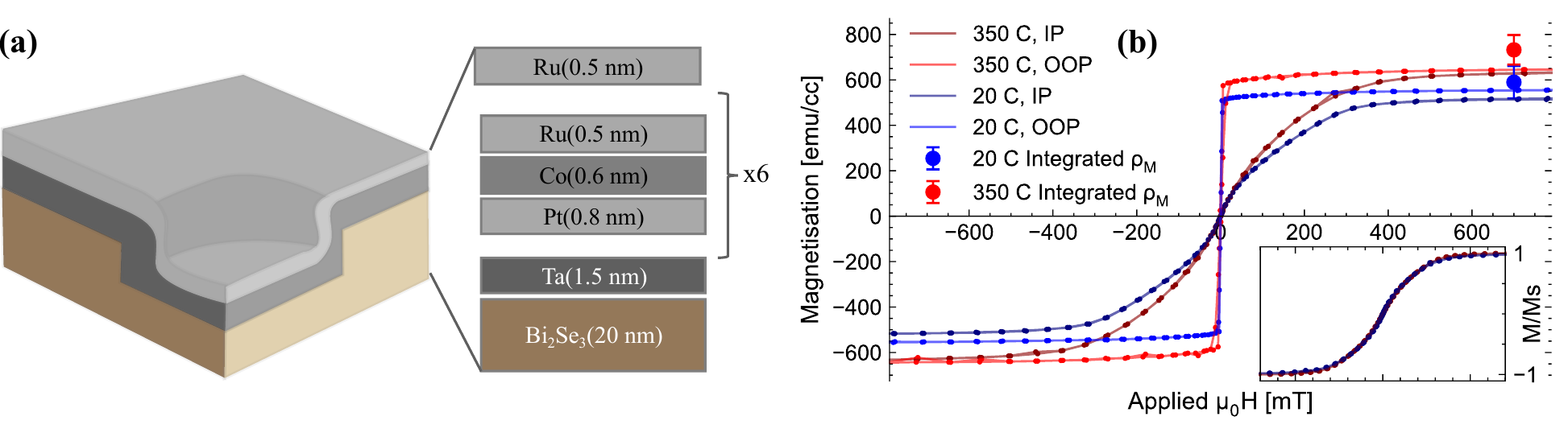}
    \caption{Sample structure and magnetic characterisation. (a) Schematic of sample including an impression of the effect of TI terracing. Here the relative scale of the terrace heights has been exaggerated for clarity; (b) SQUID VSM $M(H)$ hysteresis loops measured at room temperature in in-plane and out of plane geometries for as grown and 350\degc\ annealed samples. The magnetic moment/area value extracted from PNR slab model fittings in Fig. \ref{fig:PNR}(b) is additionally shown as the large markers. Inset axes show normalised in-plane curves for both temperatures.}
    \label{fig:intro}
\end{figure}

\begin{figure}[t]
    \centering
    \includegraphics[width=14cm]{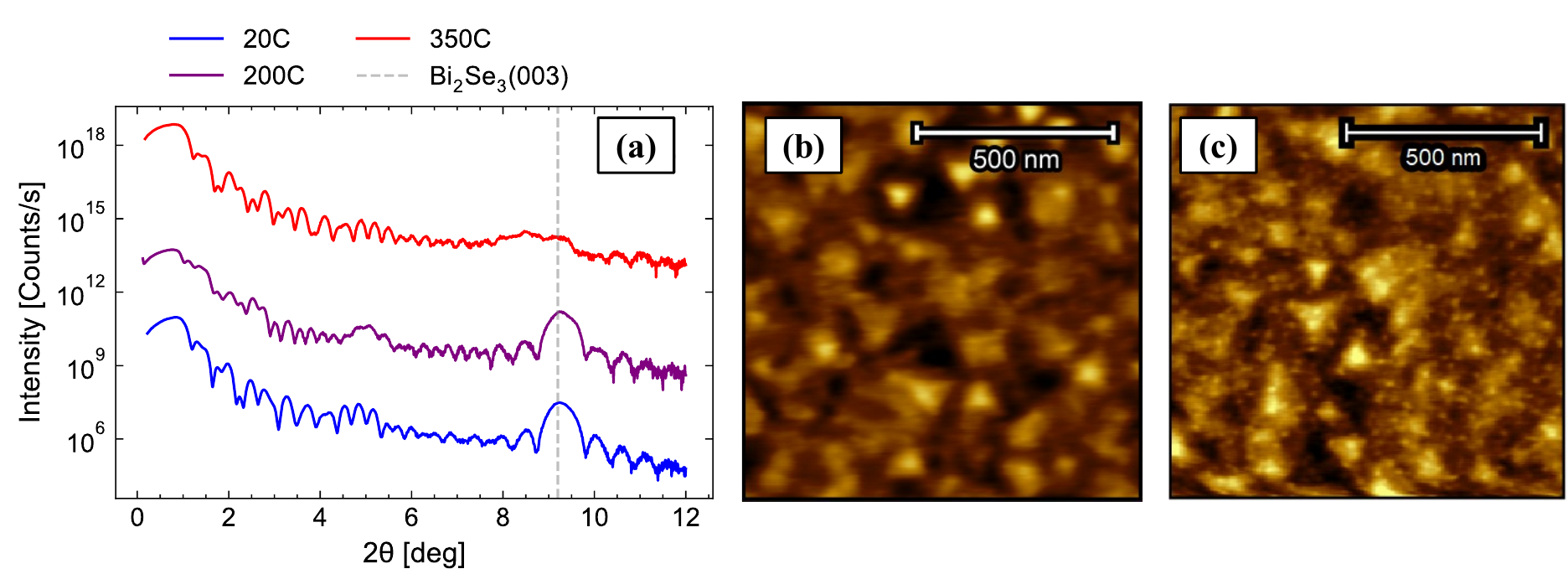}
    \caption{Structural changes to samples due to annealing.  (a) XRR curves taken after annealing at different temperatures. The \BS (003) XRD peak is marked. The curves are vertically offset for clarity; (b-c) 2 $\upmu$m $\times$ 2 $\upmu$m atomic force microscopy images of samples (b) as grown and (c) after 350\degc\ annealing.}
    \label{fig:strucuture}
\end{figure}

 Measurements of samples annealed to 350\degc\ were taken via superconducting quantum interference device vibrating sample magnetometry (SQUID-VSM) to determine any changes in magnetic properties after annealing, and are shown in Fig.~\ref{fig:intro}(b). The samples can be seen to possess strong PMA both before and after annealing. Whilst the saturation magnetisation ($M_\mathrm{s}$) was observed to increase by $\sim 20 \pm 5\%$, there was little to no change in magnetic anisotropy. Normalised in-plane hysteresis curves overlay perfectly indicating no change to the effective anisotropy field ($H_\mathrm{k}$). As we expect an increase in the demagnetisation field with magnetisation, we should expect a reduction in the effective anisotropy according to the relation:

\begin{equation}
    \mu_0 H_\mathrm{k} = \frac{1}{M_\mathrm{s}} \left( \frac{K_\mathrm{i}}{t_{\mathrm{FM}}} - \frac{1}{2}\mu_0 M_\mathrm{s}^2  \right)
\end{equation}
where $t_{\mathrm{FM}}$ is the ferromagnet thickness and $K_\mathrm{i}$ is the interfacial anisotropy energy term. The unchanged $H_\mathrm{k}$ however, suggests a corresponding increase in $K_\mathrm{i}$ to compensate. This is  consistent with an improved Co/Pt interface which explains a simultaneous increase in $M_\mathrm{s}$ and $K_\mathrm{i}$. 

X-ray reflectometry (XRR) and diffraction (XRD) measurements show that whilst the MML appears robust against high temperatures, severe breakdown to the \BS\ can take place as expected. Samples annealed to 350\degc\ display drastically lower \BS (003) XRD peak intensity as shown in XRR measurements shown in Fig.~\ref{fig:strucuture}(a). This is as well as noticeable changes to the full XRD pattern, which shows signs both of site-specific Se loss in the van der Waals gap as well as a overall loss of \BS\ crystallinity as detailed in the supplementary information \cite{supp}. Samples annealed to an intermediate temperature of 200\degc\ demonstrate much weaker changes to the XRD spectra indicating \BS\ crystallinity is much better preserved. 

As the quintuple-layer surface terracing of the TI has previously been shown to completely imprint through the multilayer \cite{Brereton2026}, the TI surface topography can hence be measured by atomic force microscopy (AFM), Fig.~\ref{fig:strucuture}(b-c). Measurements taken before and after 350\degc\ annealing show this breakdown in crystallinity. Whilst the larger terracing can still clearly be seen, small features -- likely to be of other mixed Bi-Se phases -- can be seen to have formed during the anneal process.  

\subsection{In Situ Annealing Studied by Polarised Neutron Reflectometry}

\begin{figure}[t]
    \centering
    \includegraphics[width=14cm]{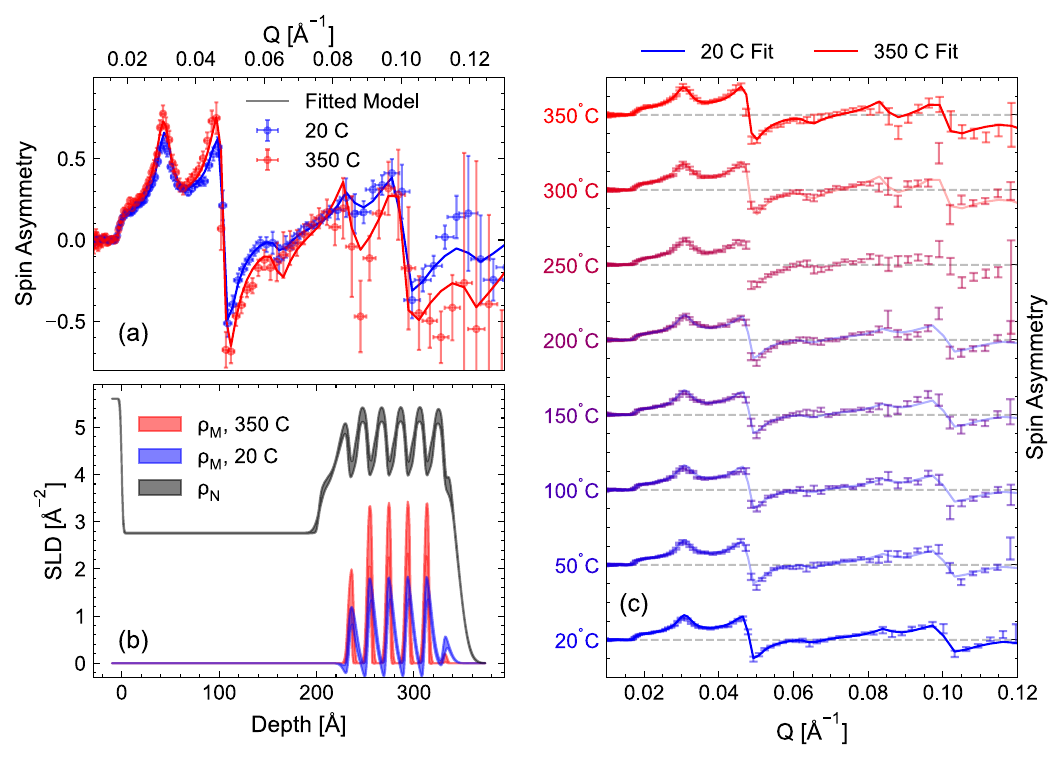}
    \caption{Polarised neutron reflectometry measurements after successive annealing steps at increasing temperature: (a) shows the spin asymmetry profiles for as-grown (20\degc\ ) and after a 350\degc\ anneal, including their respective fits corresponding to slab models for which the SLD profiles are shown in (b). Respective $2\sigma$ confidence ranges are shown. Shown in (c) are the intermediate temperature measurements with either 20 or 350\degc\ fits overlaid where they best describe the data.}
    \label{fig:PNR}
\end{figure}

With XRD and AFM measurements demonstrating significant damage sustained to the \BS\ epilayer as a result of annealing, a particular concern becomes the possibility of induced magnetic deadness in the bottom ferromagnetic layer due to the upward diffusion of Se from the TI interface. To observe this, polarised neutron reflectometry as a depth resolved magnetic probe was used. The spin-polarised neutron reflectivities $R_+$ and  $R_-$ were collected as a function of scattering wave-vector transfer $Q$ through time of flight detection. From this the spin asymmetry (SA) can be calculated, defined as $(R_{+} - R_{-})/(R_{+} + R_{-})$.

Following the shape of the SA with increasing annealing temperature, we see that this quantity remains relatively unchanged up to $\sim 250$\degc\, as shown in Fig.~\ref{fig:PNR}(c). Above this we see subtle changes at high $Q$, as well as an increase in the SA magnitude which reflects the increase in $M_\mathrm{s}$ observed in SQUID-VSM measurements. Even prior to slab modelling, due to the SA changing in shape we can deduce the magnetic profile must have changed due to annealing. 

Magnetic slab models, shown in Fig.~\ref{fig:PNR}(b), were fitted to the sample as grown and post-350\degc\ annealing with fits shown in Fig.~\ref{fig:PNR}(a). The fits were performed using the REFL1D library \cite{refl1d}. The basis of the modelling are nuclear scattering length density (SLD) $\rho_\mathrm{N}$ and magnetic SLD $\rho_\mathrm{M}$ depth profiles. We used a shared nuclear SLD profile under the assumption that changes to the multilayer reflectivity are mostly magnetic in origin -- the ML contribution to the XRR is very similar for all anneal temperatures. We additionally assume that, away from the potentially damaging TI interface or sample surface, the effect of heating should be uniform throughout the stack. Hence only the top and bottom magnetic layers are fitted independently whilst the middle four layers were modelled as identical. 
 
A trough that emerges in the SA at $Q \sim 0.09$~\AA$^{-1}$ at $T > 300$\degc , seen in Fig.~\ref{fig:PNR}(a \& c), can be shown, on examining the SLD depth profiles in Fig.~\ref{fig:PNR}(b), to correspond to the loss of magnetism in the top CoB layer and is hence independent of any breakdown of the \BS. This is instead attributed to oxidation due to the rough vacuum used and is easily avoided either by annealing under a cleaner vacuum with Ar or N, or through a thicker capping layer in contexts where this is not an issue due to imaging requirements. 

An excursion away from zero in the SA at $Q > 0.1$~\AA$^{-1}$ can be attributed to a sharpening of the magnetic interface in the magnetic slab model with the modelled magnetic interfacial roughness, $\sigma_\mathrm{M}$ collapsing from 5.0 to 0.1~\AA. Absolute values of $\sigma_\mathrm{M}$ should be treated with caution however, as here the finite range in $Q$ renders any sub-\aa ngstr\"om comparison unphysical. Importantly, the bottom layer can be seen to increase in magnetisation showing resilience against any Se diffusion. This is consistent with other studies of Bi-based TI heterostructures where other refractory metal buffer layers such as Mo and Nb have been demonstrated as effective diffusion barriers \cite{Zhang2025,Pan2022,Jalil2025}. 

Integrated $\rho_\mathrm{M}$ over $z$ can be converted into a magnetisation through  $M_\mathrm{s} = \left(\int \rho _\mathrm{M} \,\mathrm{d}z \right)/ \alpha\mu_0 t_{\text{CoB}}$ where $t_{\text{CoB}}=$ 36~\AA\ and $\alpha = 2.32$~\AA$^{-2}$T$^{-1}$ is a conversion factor \cite{Maranville2016}. Both SQUID-VSM and PNR agree on a $21 \pm 3 \%$ increase to $M_\mathrm{s}$ as averaged over the entire stack even as absolute values of magnetisation differ by $\sim 10\%$ as shown in Fig.~\ref{fig:intro}(b). In disregarding the oxidised upper layer in the slab model fit, this rises to a total $31 \pm 4\%$ increase in $M_\mathrm{s}$. 

\subsection{Effect of Thermal Treatment on Magnetic Domains}

\begin{figure}
    \centering
    \includegraphics[width=14cm]{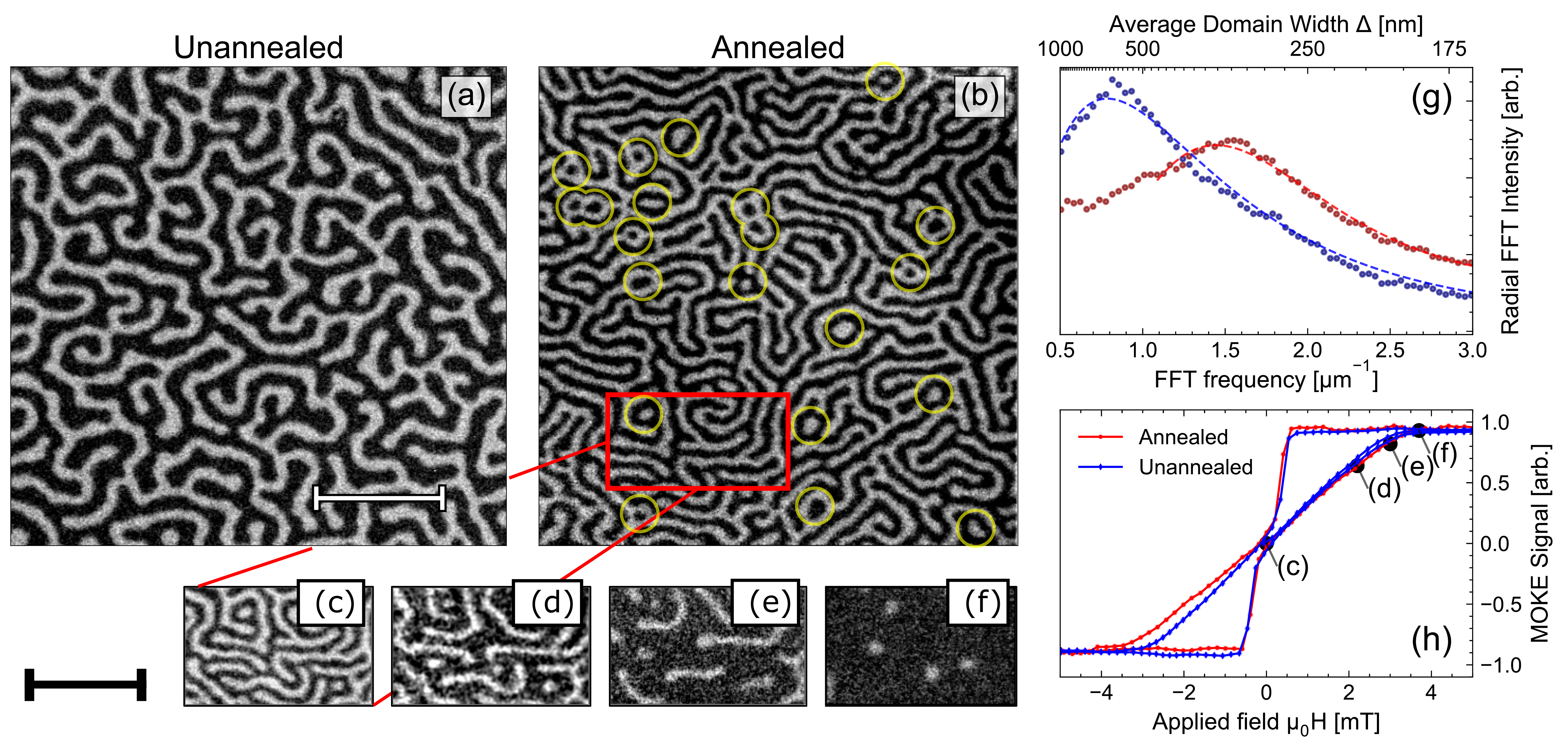}
    \caption{%Note: Is it worth making a big deal of skyrmions here? They are in zero field state so would be strange to not note but how much does it detract from main paper? 
    The effect of annealing on the observed domain structure: X-PEEM images on samples (a) before and (b) after a 200\degc\ anneal. The included scale bar in (a) represents 5 $\upmu$m for both (a) and (b). In (b) skyrmion-like features are highlighted by yellow circles. (c-f) shows field evolution of the highlighted region in (b) under an increasing applied field resulting in isolated skyrmion nucleation; (g) shows the radial FFT intensity dependence for both (a) and (b) with fitted log-normal distributions. A scale bar for (c-f) is included representing 5 $\upmu$m. (h) shows polar MOKE magnetometry under an out-of-plane applied field for both samples shown in (a) and (b). Here the corresponding fields from (c-f) are overlain.}
    \label{fig:PEEM}
\end{figure}

With both PNR and x-ray measurements showing harmful effects from annealing appearing to be minimal up to 200\degc, a sample annealed to this temperature was studied using x-ray photoemission electron microscopy (X-PEEM) with x-ray magnetic circular dichroism (XMCD) contrast. This sample shows minimal decrease to the \BS\ XRD peaks as shown in Fig.~\ref{fig:strucuture}(a) and in the supplementary information \cite{supp}. Measurements were taken at the I06 beamline at Diamond Light Source where incident x-rays were tuned to the Co L$_3$ absorption edge. Images of both photon helicities were taken, normalised and subtracted to observe changes to the magnetic domain structure.

As shown in Fig. \ref{fig:PEEM}(a-b), changes to the zero field domains are apparent. The average domain width, shown in Fig.~\ref{fig:PEEM}(g), extracted from angular averages of fast Fourier transforms {FFTs} of the images, was reduced by $\sim 40\%$ from 600~to 340 nm and becomes more sharply defined with the standard deviation in the fitted log-normal distribution of the FFT peak narrowing by 20\% from 1.9 to 1.5 \um$^{-1}$. Little to no change is seen in the out-of-plane hysteresis suggesting many functional properties of the MML remain unchanged as shown in \ref{fig:PEEM}(h) by polar magneto-optical Kerr effect (MOKE) magnetometry. This appears to be consistent with the unchanged $H_{\mathrm{k}}$ observed after annealing by in-plane SQUID-VSM. 

Also seen at zero field are new skyrmion-like features. These are seen to persist to higher field than the maze domains when an external, out-of-plane field is applied, as shown in Fig. \ref{fig:PEEM}(c-f). This relative stability against permutations supports the assumption they are topological. Skyrmions could not similarly be nucleated through magnetic field in the control sample reinforcing the need for prior annealing of the MML. This is similar to studies of Pt/Co-based multilayers showing that thermal treatment aids skyrmion nucleation \cite{Zhang2018, Wang2021}. It is likely both the improved domain ordering and skyrmion stabilisation are due to an increase in the strength of the interfacial DMI. This would be consistent with the  sharpening of the magnetic interfaces seen in PNR slab model fitting \cite{Wells2017} and the corresponding increase in $K_i$ from SQUID-VSM magnetometry. 

\section{Conclusions}

These samples have proven to be structurally and magnetically robust up to at least 200\degc\, in agreement with wider literature \cite{Zhang2025, Pan2022, Bansal2011, He2011, Puthran2024, Sapkota2021}. This will be a satisfactory range for reliable operation in the context of spintronic devices for most applications. Interdiffusion between the TI and multilayer is effectively suppressed at the interface by the Ta buffer layer given the lack of any magnetic deadness appearing during annealing. Device function under increasing temperature instead appears to be limited by the breakdown in the \BS\ structure driven by Se loss, not the multilayer. The known efficacy of refractory buffer layers as a diffusion barrier against Se \cite{Zhang2025,Pan2022,Jalil2025} is further established and marks the first time this has been demonstrated by PNR. %This may not be a hard limit however, as other studies describe annealing processes in the range of $300 - 450$\degc\ for \BX-metal films \cite{Pan2022}. Whilst here changes to TI crystallinity may be of less concern, these studies demonstrate no loss of TI functionality as a result. %The result obtained in this work is consistent with previous studies where $\sim 250$\degc\ is found to be the effective upper limit of \BX\ device function \cite{Shu2025, Sapkota2021, Lin2011}.  

The existence of skyrmions in a TI-multilayer heterostructure is promising for future TSS-driven skyrmion motion experiments. Whilst previous studies have shown skyrmions or a corresponding topological Hall effect hosted in other TI-based heterostructures \cite{chen2019, Wu2020}, this marks their first documented observation in a heterostructure that combines a traditional ferromagnet - heavy metal multilayer system with a TI. Such multilayer systems are known for being widely customisable through the control of many different growth parameters and may hence prove more versatile in their integration with topologically insulating materials. This is in addition to the physics of SOT-driven spin texture manipulation in chiral multilayers being, at this stage of the field, widely documented and well understood \cite{Woo2016,Hamamoto2015,Romming2013,Jiang2017,Zeissler2020, Buttner2017}. In taking advantage of this well established literature, we show that the annealing processes often used in conventional, domain-hosting, Pt/Co-based films, can be successfully transferred to the TI-MML heterostructures to preferentially alter magnetic properties: improving the zero field domain state and aiding skyrmion nucleation.

\ack{The experiment at the ISIS Neutron and Muon Source was supported by beamtime allocation RB2500062 from the Science and Technology Facilities Council. XMCD-PEEM imaging was undertaken at Diamond Light Source under sessions MM37770-3 and MM38770-1.} %Acknowledgment is given to Dirk Backes at Diamond Light Source who has helped supervise this project since these experiments were undertaken.}

\funding{S.H., T.A.M. and C.H.M. acknowledge support from the EPSRC Programme Grant `CAMIE' (EP/X027074/1), A.Y. and S.S. also acknowledge `NAME' (EP/V001914/1).
The samples were grown in the Royce Deposition system at the University of Leeds, which is supported by the Henry Royce Institute, United Kingdom, through Grants No. EP/P022464/1 and No. EP/R00661X/1. }
% This section is a list of funder names and grant numbers

%\roles{}
% List author names and the contributions made to the article, using terms from the NISO Contributor Roles Taxonomy (CRediT) https://credit.niso.org

\data{The data associated with this paper are openly available at the Research Data Leeds repository~\cite{data}. Raw data associated with the ISIS experiment is available at the URL: \url{https://doi.org/10.5286/ISIS.E.RB2500062}}
% For more information on IOP Publishing's research data policy see: https://publishingsupport.iopscience.iop.org/questions/research-data/

%\suppdata{Sample text inserted for demonstration.}

%\section*{References}
\bibliographystyle{iopart-num}

\bibliography{library.bib}

@article{Wells2017,
  title = {Effect of interfacial intermixing on the Dzyaloshinskii-Moriya interaction in Pt/Co/Pt},
  author = {Wells, Adam W. J. and Shepley, Philippa M. and Marrows, Christopher H. and Moore, Thomas A.},
  journal = {Phys. Rev. B},
  volume = {95},
  pages = {054428},
  year = {2017},
  doi = {10.1103/PhysRevB.95.054428}
}

@article{Wang2017,
   author = {Yi Wang and Dapeng Zhu and Yang Wu and Yumeng Yang and Jiawei Yu and Rajagopalan Ramaswamy and Rahul Mishra and Shuyuan Shi and Mehrdad Elyasi and Kie Leong Teo and Yihong Wu and Hyunsoo Yang},
   doi = {10.1038/s41467-017-01583-4},
   IGNORE = {20411723},
   IGNORE = {1},
   journal = {Nat. Commun.},
   IGNORE = {12},
   pmid = {29118331},
   publisher = {Nature Publishing Group},
   title = {Room temperature magnetization switching in topological insulator-ferromagnet heterostructures by spin-orbit torques},
   volume = {8},
   year = {2017},
   pages = {1364}
}

@article{Zeissler2020,
   author = {Katharina Zeissler and Simone Finizio and Craig Barton and Alexandra J. Huxtable and Jamie Massey and Jörg Raabe and Alexandr V. Sadovnikov and Sergey A. Nikitov and Richard Brearton and Thorsten Hesjedal and Gerrit van der Laan and Mark C. Rosamond and Edmund H. Linfield and Gavin Burnell and Christopher H. Marrows},
   doi = {10.1038/s41467-019-14232-9},
   IGNORE = {20411723},
   IGNORE = {1},
   journal = {Nat. Commun.},
   IGNORE = {12},
   pmid = {31969569},
   publisher = {Nature Research},
   title = {Diameter-independent skyrmion Hall angle observed in chiral magnetic multilayers},
   volume = {11},
   year = {2020},
   pages = {428}
}

@article{Heinze2011,
   author = {Stefan Heinze and Kirsten Von Bergmann and Matthias Menzel and Jens Brede and André Kubetzka and Roland Wiesendanger and Gustav Bihlmayer and Stefan Blügel},
   doi = {10.1038/nphys2045},
   IGNORE = {17452481},
   IGNORE = {9},
   journal = {Nat. Phys.},
   pages = {713-718},
   publisher = {Nature Publishing Group},
   title = {Spontaneous atomic-scale magnetic skyrmion lattice in two dimensions},
   volume = {7},
   year = {2011},
}

@article{Moreau,
   author = {C. Moreau-Luchaire and C. Moutafis and N. Reyren and J. Sampaio and C. A.F. Vaz and N. Van Horne and K. Bouzehouane and K. Garcia and C. Deranlot and P. Warnicke and P. Wohlhüter and J. M. George and M. Weigand and J. Raabe and V. Cros and A. Fert},
   doi = {10.1038/nnano.2015.313},
   IGNORE = {17483395},
   IGNORE = {5},
   journal = {Nat. Nanotechnol.},
   IGNORE = {5},
   pages = {444-448},
   pmid = {26780660},
   publisher = {Nature Publishing Group},
   title = {Additive interfacial chiral interaction in multilayers for stabilization of small individual skyrmions at room temperature},
   volume = {11},
   year = {2016},
}

@article{Woo2016,
   author = {Seonghoon Woo and Kai Litzius and Benjamin Krüger and Mi Young Im and Lucas Caretta and Kornel Richter and Maxwell Mann and Andrea Krone and Robert M. Reeve and Markus Weigand and Parnika Agrawal and Ivan Lemesh and Mohamad Assaad Mawass and Peter Fischer and Mathias Kläui and Geoffrey S.D. Beach},
   doi = {10.1038/nmat4593},
   IGNORE = {14764660},
   IGNORE = {5},
   journal = {Nat. Mater.},
   IGNORE = {5},
   pages = {501-506},
   pmid = {26928640},
   publisher = {Nature Publishing Group},
   title = {Observation of room-temperature magnetic skyrmions and their current-driven dynamics in ultrathin metallic ferromagnets},
   volume = {15},
   year = {2016},
}

@article{Han2017,
   author = {Jiahao Han and A. Richardella and Saima A. Siddiqui and Joseph Finley and N. Samarth and Luqiao Liu},
   doi = {10.1103/PhysRevLett.119.077702},
   IGNORE = {0031-9007},
   IGNORE = {7},
   journal = {Phys. Rev. Lett.},
   IGNORE = {8},
   pages = {077702},
   title = {Room-Temperature Spin-Orbit Torque Switching Induced by a Topological Insulator},
   volume = {119},
   year = {2017},
}

@article{Wu2020,
   author = {Hao Wu and Felix Groß and Bingqian Dai and David Lujan and Seyed Armin Razavi and Peng Zhang and Yuxiang Liu and Kemal Sobotkiewich and Johannes Förster and Markus Weigand and Gisela Schütz and Xiaoqin Li and Joachim Gräfe and Kang L. Wang},
   doi = {10.1002/adma.202003380},
   IGNORE = {0935-9648},
   IGNORE = {34},
   journal = {Adv. Mater.},
   IGNORE = {8},
   title = {Ferrimagnetic Skyrmions in Topological Insulator/Ferrimagnet Heterostructures},
   volume = {32},
   year = {2020},
   pages = {2003380}
}

@article{Chen2019,
   author = {Junshu Chen and Linjing Wang and Meng Zhang and Liang Zhou and Runnan Zhang and Lipeng Jin and Xuesen Wang and Hailang Qin and Yang Qiu and Jiawei Mei and Fei Ye and Bin Xi and Hongtao He and Bin Li and Gan Wang},
   doi = {10.1021/acs.nanolett.9b02191},
   IGNORE = {15306992},
   IGNORE = {9},
   journal = {Nano Letters},
   IGNORE = {9},
   pages = {6144-6151},
   pmid = {31438678},
   publisher = {American Chemical Society},
   title = {Evidence for Magnetic Skyrmions at the Interface of Ferromagnet/Topological-Insulator Heterostructures},
   volume = {19},
   year = {2019},
}

@article{Fan2014,
   author = {Yabin Fan and Pramey Upadhyaya and Xufeng Kou and Murong Lang and So Takei and Zhenxing Wang and Jianshi Tang and Liang He and Li Te Chang and Mohammad Montazeri and Guoqiang Yu and Wanjun Jiang and Tianxiao Nie and Robert N. Schwartz and Yaroslav Tserkovnyak and Kang L. Wang},
   doi = {10.1038/nmat3973},
   IGNORE = {14764660},
   IGNORE = {7},
   journal = {Nat. Mater.},
   pages = {699-704},
   publisher = {Nature Publishing Group},
   title = {Magnetization switching through giant spin-orbit torque in a magnetically doped topological insulator heterostructure},
   volume = {13},
   year = {2014},
}

@article{Dc2018,
   author = {Mahendra Dc and Roberto Grassi and Jun Yang Chen and Mahdi Jamali and Danielle Reifsnyder Hickey and Delin Zhang and Zhengyang Zhao and Hongshi Li and P. Quarterman and Yang Lv and Mo Li and Aurelien Manchon and K. Andre Mkhoyan and Tony Low and Jian Ping Wang},
   doi = {10.1038/s41563-018-0136-z},
   IGNORE = {14764660},
   IGNORE = {9},
   journal = {Nat. Mater.},
   IGNORE = {9},
   pages = {800-807},
   pmid = {30061733},
   publisher = {Nature Publishing Group},
   title = {Room-temperature high spin–orbit torque due to quantum confinement in sputtered BixSe(1–x) films},
   volume = {17},
   year = {2018},
}

@article{Romming2013,
   author = {Niklas Romming and Christian Hanneken and Matthias Menzel and Jessica E. Bickel and Boris Wolter and Kirsten von Bergmann and André Kubetzka and Roland Wiesendanger},
   doi = {10.1126/science.1240573},
   IGNORE = {0036-8075},
   IGNORE = {6146},
   journal = {Science},
   IGNORE = {8},
   pages = {636-639},
   title = {Writing and Deleting Single Magnetic Skyrmions},
   volume = {341},
   year = {2013},
}

@article{Zhang2009,
   author = {Haijun Zhang and Chao-Xing Liu and Xiao-Liang Qi and Xi Dai and Zhong Fang and Shou-Cheng Zhang},
   doi = {10.1038/nphys1270},
   IGNORE = {1745-2473},
   IGNORE = {6},
   journal = {Nat. Phys.},
   IGNORE = {6},
   pages = {438-442},
   title = {Topological insulators in Bi2Se3, Bi2Te3 and Sb2Te3 with a single Dirac cone on the surface},
   volume = {5},
   year = {2009},
}

@article{Li2014,
   author = {C. H. Li and O. M. J. van ‘t Erve and J. T. Robinson and Y. Liu and L. Li and B. T. Jonker},
   doi = {10.1038/nnano.2014.16},
   IGNORE = {1748-3387},
   IGNORE = {3},
   journal = {Nat. Nanotechnol.},
   IGNORE = {3},
   pages = {218-224},
   title = {Electrical detection of charge-current-induced spin polarization due to spin-momentum locking in Bi2Se3},
   volume = {9},
   year = {2014},
}

@article{Shiomi2014,
   author = {Y. Shiomi and K. Nomura and Y. Kajiwara and K. Eto and M. Novak and Kouji Segawa and Yoichi Ando and E. Saitoh},
   doi = {10.1103/PhysRevLett.113.196601},
   IGNORE = {0031-9007},
   IGNORE = {19},
   journal = {Phys. Rev. Lett.},
   IGNORE = {11},
   pages = {196601},
   title = {Spin-Electricity Conversion Induced by Spin Injection into Topological Insulators},
   volume = {113},
   year = {2014},
}

@article{Xia2009,
   author = {Y. Xia and D. Qian and D. Hsieh and L. Wray and A. Pal and H. Lin and A. Bansil and D. Grauer and Y. S. Hor and R. J. Cava and M. Z. Hasan},
   doi = {10.1038/nphys1274},
   IGNORE = {1745-2473},
   IGNORE = {6},
   journal = {Nat. Phys.},
   IGNORE = {6},
   pages = {398-402},
   title = {Observation of a large-gap topological-insulator class with a single Dirac cone on the surface},
   volume = {5},
   year = {2009},
}

@article{Jiang2017,
   author = {Wanjun Jiang and Xichao Zhang and Guoqiang Yu and Wei Zhang and Xiao Wang and M. Benjamin Jungfleisch and John E. Pearson and Xuemei Cheng and Olle Heinonen and Kang L. Wang and Yan Zhou and Axel Hoffmann and Suzanne G. E. te Velthuis},
   doi = {10.1038/nphys3883},
   IGNORE = {1745-2473},
   IGNORE = {2},
   journal = {Nat. Phys.},
   IGNORE = {2},
   pages = {162-169},
   title = {Direct observation of the skyrmion Hall effect},
   volume = {13},
   year = {2017},
}

@article{Pan2022,
   author = {Quanjun Pan and Yuting Liu and Hao Wu and Peng Zhang and Hanshen Huang and Christopher Eckberg and Xiaoyu Che and Yingying Wu and Bingqian Dai and Qiming Shao and Kang L. Wang},
   doi = {10.1002/aelm.202200003},
   IGNORE = {2199-160X},
   IGNORE = {9},
   journal = {Adv. Electron. Mater.},
   IGNORE = {9},
   title = {Efficient Spin-Orbit Torque Switching of Perpendicular Magnetization using Topological Insulators with High Thermal Tolerance},
   volume = {8},
   year = {2022},
   pages = {2200003}
}

@article{Mellnik2014,
   author = {A. R. Mellnik and J. S. Lee and A. Richardella and J. L. Grab and P. J. Mintun and M. H. Fischer and A. Vaezi and A. Manchon and E.-A. Kim and N. Samarth and D. C. Ralph},
   doi = {10.1038/nature13534},
   IGNORE = {0028-0836},
   IGNORE = {7510},
   journal = {Nature},
   IGNORE = {7},
   pages = {449-451},
   title = {Spin-transfer torque generated by a topological insulator},
   volume = {511},
   year = {2014},
}

@article{Zhang2025,
   author = {Xu Zhang and Aitian Chen and Yifan Zhang and Zhaozhuo Zeng and Yaqin Guo and Dongxing Zheng and Baoshan Cui and Chuangwen Wu and Wenjie Song and Shuo Yang and Zijun Luo and Jingfeng Li and Gianluca Gubbiotti and Xiufeng Han and Jinkui Zhao and Peng Yan and Xufeng Kou and Xixiang Zhang and Hao Wu},
   doi = {10.1002/aelm.202500022},
   IGNORE = {2199-160X},
   journal = {Adv. Electron. Mater.},
   IGNORE = {4},
   title = {Ultralow Electrical Current Driven Field‐Free Spin‐Orbit Torque Switching of Magnetic Tunnel Junctions by Topological Insulators},
   year = {2025},
   volume = {11},
   pages = {2500022}
}

@article{Fan2016,
   author = {Yabin Fan and Xufeng Kou and Pramey Upadhyaya and Qiming Shao and Lei Pan and Murong Lang and Xiaoyu Che and Jianshi Tang and Mohammad Montazeri and Koichi Murata and Li-Te Chang and Mustafa Akyol and Guoqiang Yu and Tianxiao Nie and Kin L. Wong and Jun Liu and Yong Wang and Yaroslav Tserkovnyak and Kang L. Wang},
   doi = {10.1038/nnano.2015.294},
   IGNORE = {1748-3387},
   IGNORE = {4},
   journal = {Nat. Nanotechnol.},
   IGNORE = {4},
   pages = {352-359},
   title = {Electric-field control of spin–orbit torque in a magnetically doped topological insulator},
   volume = {11},
   year = {2016},
}

@article{Maranville2016,
   author = {Brian B. Maranville and Brian J. Kirby and Alexander J. Grutter and Paul A. Kienzle and Charles F. Majkrzak and Yaohua Liu and Cindi L. Dennis},
   doi = {10.1107/S1600576716007135},
   IGNORE = {1600-5767},
   IGNORE = {4},
   journal = {J. Appl. Crystallogr.},
   IGNORE = {8},
   pages = {1121-1129},
   title = {Measurement and modeling of polarized specular neutron reflectivity in large magnetic fields},
   volume = {49},
   year = {2016},
}

@article{Hamamoto2015,
   author = {Keita Hamamoto and Motohiko Ezawa and Naoto Nagaosa},
   doi = {10.1103/PhysRevB.92.115417},
   IGNORE = {1098-0121},
   IGNORE = {11},
   journal = {Phys. Rev. B},
   IGNORE = {9},
   pages = {115417},
   title = {Quantized topological Hall effect in skyrmion crystal},
   volume = {92},
   year = {2015},
}

@misc{refl1d,
   author = {Brian Benjamin Maranville and Paul Kienzle and Richard Sheridan and Andrew J. Caruana and Jose Borreguero and Andrew Nelson and David P. Hoogerheide and Alexander Book and Marie Blackman and Purnima P. Balakrishnan and Ronnie Ghose},
   doi = {https://dx.doi.org/10.5281/zenodo.1249715},
   url = {https://github.com/reflectometry/refl1d/tree/v1.0.2b1},
   title = {reflectometry/refl1d: v1.0.0},
   
   year = {2025},
}

@article{Chen2015,
   author = {Gong Chen and Arantzazu Mascaraque and Alpha T. N'Diaye and Andreas K. Schmid},
   doi = {10.1063/1.4922726},
   IGNORE = {0003-6951},
   IGNORE = {24},
   journal = {Appl. Phys. Lett.},
   IGNORE = {6},
   title = {Room temperature skyrmion ground state stabilized through interlayer exchange coupling},
   volume = {106},
   year = {2015},
   pages = {242404}
}

@article{Jiang2015,
   author = {Wanjun Jiang and Pramey Upadhyaya and Wei Zhang and Guoqiang Yu and M. Benjamin Jungfleisch and Frank Y. Fradin and John E. Pearson and Yaroslav Tserkovnyak and Kang L. Wang and Olle Heinonen and Suzanne G. E. te Velthuis and Axel Hoffmann},
   doi = {10.1126/science.aaa1442},
   IGNORE = {0036-8075},
   IGNORE = {6245},
   journal = {Science},
   IGNORE = {7},
   pages = {283-286},
   title = {Blowing magnetic skyrmion bubbles},
   volume = {349},
   year = {2015},
}

@article{Soumyanarayanan2017,
   author = {Anjan Soumyanarayanan and M. Raju and A. L. Gonzalez Oyarce and Anthony K. C. Tan and Mi-Young Im and A. P. Petrović and Pin Ho and K. H. Khoo and M. Tran and C. K. Gan and F. Ernult and C. Panagopoulos},
   doi = {10.1038/nmat4934},
   IGNORE = {1476-1122},
   IGNORE = {9},
   journal = {Nat. Mater.},
   IGNORE = {9},
   pages = {898-904},
   title = {Tunable room-temperature magnetic skyrmions in Ir/Fe/Co/Pt multilayers},
   volume = {16},
   year = {2017},
}

@article{Buttner2017,
   author = {Felix Büttner and Ivan Lemesh and Michael Schneider and Bastian Pfau and Christian M. Günther and Piet Hessing and Jan Geilhufe and Lucas Caretta and Dieter Engel and Benjamin Krüger and Jens Viefhaus and Stefan Eisebitt and Geoffrey S. D. Beach},
   doi = {10.1038/nnano.2017.178},
   IGNORE = {1748-3387},
   IGNORE = {11},
   journal = {Nat. Nanotechnol.},
   IGNORE = {11},
   pages = {1040-1044},
   title = {Field-free deterministic ultrafast creation of magnetic skyrmions by spin–orbit torques},
   volume = {12},
   year = {2017},
}

@article{He2011,
   author = {Liang He and Faxian Xiu and Yong Wang and Alexei V. Fedorov and Guan Huang and Xufeng Kou and Murong Lang and Ward P. Beyermann and Jin Zou and Kang L. Wang},
   doi = {10.1063/1.3585673},
   IGNORE = {0021-8979},
   IGNORE = {10},
   journal = {J. Appl. Phys.},
   IGNORE = {5},
   title = {Epitaxial growth of Bi2Se3 topological insulator thin films on Si (111)},
   volume = {109},
   year = {2011},
   pages = {103702}
}

@article{Bansal2011,
   author = {Namrata Bansal and Yong Seung Kim and Eliav Edrey and Matthew Brahlek and Yoichi Horibe and Keiko Iida and Makoto Tanimura and Guo-Hong Li and Tian Feng and Hang-Dong Lee and Torgny Gustafsson and Eva Andrei and Seongshik Oh},
   doi = {10.1016/j.tsf.2011.07.033},
   IGNORE = {00406090},
   IGNORE = {1},
   journal = {Thin Solid Films},
   IGNORE = {10},
   pages = {224-229},
   title = {Epitaxial growth of topological insulator Bi2Se3 film on Si(111) with atomically sharp interface},
   volume = {520},
   year = {2011},
}

@article{Brereton2026,
   author = {Benjamin A. Brereton and Soumyarup Hait and Ahmet Yagmur and Christy J. Kinane and Francesco Maccherozzi and Michele Conroy and Satoshi Sasaki and Thomas A. Moore and Sarnjeet S. Dhesi and Sean Langridge and Christopher H. Marrows},
   doi = {10.1103/hp66-tldh},
   IGNORE = {2475-9953},
   IGNORE = {6},
   journal = {Phys. Rev. Mater.},
   IGNORE = {6},
   pages = {064413},
   title = {Tailoring ultrathin magnetic multilayers at terraced topologically insulating interfaces for perpendicularly magnetized domains},
   volume = {10},
   year = {2026},
}

@article{Puthran2024,
   author = {Suchitra Puthran and Ganesh Shridhar Hegde and A. N. Prabhu and Vikash Mishra and Tzu-Yi Yang and Y. K. Kuo},
   doi = {10.1007/s10853-024-09865-1},
   IGNORE = {0022-2461},
   IGNORE = {26},
   journal = {J. Mater. Sci.},
   IGNORE = {7},
   pages = {11817-11834},
   title = {A comprehensive experimental and DFT analysis on thermoelectric properties of Sb, Te co-doped Bi2Se3 polycrystals},
   volume = {59},
   year = {2024},
}

@article{Bayaz2003,
   author = {A. Al Bayaz and A. Giani and A. Foucaran and F. Pascal-Delannoy and A. Boyer},
   doi = {10.1016/S0040-6090(03)00675-8},
   IGNORE = {00406090},
   IGNORE = {1-2},
   journal = {Thin Solid Films},
   IGNORE = {9},
   pages = {1-5},
   title = {Electrical and thermoelectrical properties of Bi2Se3 grown by metal organic chemical vapour deposition technique},
   volume = {441},
   year = {2003},
}

@article{Mishra1997,
   author = {S K Mishra and S Satpathy and O Jepsen},
   doi = {10.1088/0953-8984/9/2/014},
   IGNORE = {0953-8984},
   IGNORE = {2},
   journal = {J. Phys.: Condens. Matter},
   IGNORE = {1},
   pages = {461-470},
   title = {Electronic structure and thermoelectric properties of bismuth telluride and bismuth selenide},
   volume = {9},
   year = {1997},
}

@article{Kadel2010,
   author = {K Kadel and Latha Kumari and WZ Li and Jian Yu Huang and Paula P Provencio},
   doi = {10.1007/s11671-010-9795-7},
   IGNORE = {1556-276X},
   IGNORE = {1},
   journal = {Nanoscale Res. Lett.},
   IGNORE = {10},
   pages = {57},
   title = {Synthesis and Thermoelectric Properties of Bi2Se3 Nanostructures},
   volume = {6},
   year = {2010},
}

@article{Sapkota2021,
   author = {Yub Raj Sapkota and Dipanjan Mazumdar},
   doi = {10.1016/j.tsf.2021.138676},
   IGNORE = {00406090},
   journal = {Thin Solid Films},
   IGNORE = {6},
   pages = {138676},
   title = {Influence of post-deposition annealing on the transport properties of sputtered Bi2Se3 thin films},
   volume = {727},
   year = {2021},
}

@article{Hait2026,
   author = {Soumyarup Hait and Benjamin A. Brereton and Ahmet Yagmur and Satoshi Sasaki and Gavin Burnell and Christopher H. Marrows},
   IGNORE = {8},
   title = {Temperature-Dependent Spin-Orbit Torque Generation in Perpendicularly Magnetized Topological Insulator-Magnetic Multilayer Heterostructures},
   year = {2026},
   journal = {cond-mat.mes-hall},
   eprint = {arXiv:2608.15232},
   primaryClass = {cond-mat.mes-hall},
}

@article{Cafaro1984,
   author = {Marcella Lucia Cafaro and Gianpiero Bardi and Vincenzo Piacente},
   doi = {10.1021/je00035a026},
   IGNORE = {0021-9568},
   IGNORE = {1},
   journal = {Journal of Chemical \& Engineering Data},
   IGNORE = {1},
   pages = {78-80},
   title = {Vaporization study of solid bismuth selenide (Bi2Se3)},
   volume = {29},
   year = {1984},
}

@article{Stull1947,
   author = {Daniel R. Stull},
   doi = {10.1021/ie50448a022},
   IGNORE = {0019-7866},
   IGNORE = {4},
   journal = {Industrial \& Engineering Chemistry},
   IGNORE = {4},
   pages = {517-550},
   title = {Vapor Pressure of Pure Substances. Organic and Inorganic Compounds},
   volume = {39},
   year = {1947},
}

@article{Momma2011,
   author = {Koichi Momma and Fujio Izumi},
   doi = {10.1107/S0021889811038970},
   IGNORE = {0021-8898},
   IGNORE = {6},
   journal = {J. Appl. Crystallogr.},
   IGNORE = {12},
   pages = {1272-1276},
   title = {<i>VESTA 3</i> for three-dimensional visualization of crystal, volumetric and morphology data},
   volume = {44},
   year = {2011},
}

@article{Xue2013,
   author = {L. Xue and P. Zhou and C. X. Zhang and C. Y. He and G. L. Hao and L. Z. Sun and J. X. Zhong},
   doi = {10.1063/1.4804439},
   IGNORE = {2158-3226},
   IGNORE = {5},
   journal = {AIP Advances},
   IGNORE = {5},
   title = {First-principles study of native point defects in Bi2Se3},
   volume = {3},
   year = {2013},
   pages = {052105}
}

@article{Jalil2025,
   author = {Abdur Rehman Jalil and Tobias W. Schmitt and Philipp Rüßmann and Xian‐Kui Wei and Benedikt Frohn and Michael Schleenvoigt and Wilhelm Wittl and Xiao Hou and Anne Schmidt and Kaycee Underwood and Gustav Bihlmayer and Martina Luysberg and Joachim Mayer and Stefan Blügel and Detlev Grützmacher and Peter Schüffelgen},
   doi = {10.1002/qute.202400343},
   IGNORE = {2511-9044},
   IGNORE = {3},
   journal = {Adv. Quantum Technol.},
   IGNORE = {3},
   title = {Engineering Epitaxial Interfaces for Topological Insulator — Superconductor Hybrid Devices with Al Electrodes},
   volume = {8},
   year = {2025},
   pages = {2400343}
}

@article{Wang2021,
   author = {Xinran Wang and Anni Cao and Sai Li and Jin Tang and Ao Du and Houyi Cheng and Yiming Sun and Haifeng Du and Xueying Zhang and Weisheng Zhao},
   doi = {10.1103/PhysRevB.104.064421},
   IGNORE = {2469-9950},
   IGNORE = {6},
   journal = {Phys. Rev. B},
   IGNORE = {8},
   pages = {064421},
   title = {Manipulating density of magnetic skyrmions via multilayer repetition and thermal annealing},
   volume = {104},
   year = {2021},
}

@article{Rodmacq2009,
   author = {B. Rodmacq and A. Manchon and C. Ducruet and S. Auffret and B. Dieny},
   doi = {10.1103/PhysRevB.79.024423},
   IGNORE = {1098-0121},
   IGNORE = {2},
   journal = {Phys. Rev. B},
   IGNORE = {1},
   pages = {024423},
   title = {Influence of thermal annealing on the perpendicular magnetic anisotropy of Pt/Co/AlOx trilayers},
   volume = {79},
   year = {2009},
}

@article{Bandiera2012,
   author = {S. Bandiera and R. C. Sousa and B. Rodmacq and B. Dieny},
   doi = {10.1063/1.3701585},
   IGNORE = {0003-6951},
   IGNORE = {14},
   journal = {Appl. Phys. Lett.},
   IGNORE = {4},
   title = {Enhancement of perpendicular magnetic anisotropy through reduction of Co-Pt interdiffusion in (Co/Pt) multilayers},
   volume = {100},
   year = {2012},
   pages = {142410}
}

@article{Cao2020,
   author = {Anni Cao and Runze Chen and Xinran Wang and Xueying Zhang and Shiyang Lu and Shishen Yan and Bert Koopmans and Weisheng Zhao},
   doi = {10.1088/1361-6528/ab62cd},
   IGNORE = {0957-4484},
   IGNORE = {15},
   journal = {Nanotechnology},
   IGNORE = {4},
   pages = {155705},
   title = {Enhanced interfacial Dzyaloshinskii—Moriya interactions in annealed Pt/Co/MgO structures},
   volume = {31},
   year = {2020},
}

@article{Zhang2018,
   author = {Senfu Zhang and Junwei Zhang and Yan Wen and Eugene M. Chudnovsky and Xixiang Zhang},
   doi = {10.1063/1.5053983},
   IGNORE = {0003-6951},
   IGNORE = {19},
   journal = {Appl. Phys. Lett.},
   IGNORE = {11},
   title = {Creation of a thermally assisted skyrmion lattice in Pt/Co/Ta multilayer films},
   volume = {113},
   year = {2018},
   pages = {192403}
}

@article{Gweon2018,
   author = {Hyung Keun Gweon and Sang Ho Lim},
   doi = {10.7567/JJAP.57.030301},
   IGNORE = {0021-4922},
   IGNORE = {3},
   journal = {Jpn. J. Appl. Phys.},
   IGNORE = {3},
   pages = {030301},
   title = {Relative strength of perpendicular magnetic anisotropy at bottom and top interfaces in Pt/Co/MgO trilayers},
   volume = {57},
   year = {2018},
}

@article{Kim2019,
   author = {Woo-Yeong Kim and Hyung Keun Gweon and Kyung-Jin Lee and Chun-Yeol You},
   doi = {10.7567/1882-0786/ab181c},
   IGNORE = {1882-0778},
   IGNORE = {5},
   journal = {Appl. Phys. Express},
   IGNORE = {5},
   pages = {053007},
   title = {Correlation between interfacial Dzyaloshinskii–Moriya interaction and interfacial magnetic anisotropy of Pt/Co/MgO structures},
   volume = {12},
   year = {2019},
}

\end{document}

\maketitle

\section{Measurement Details}
\subsection{SQUID-VSM}

Measurements were conducted on a Quantum Design MPMS-3 SQUID-VSM magnetometer and were measured over a $\pm 2$ T range at room temperature. These were subsequently corrected for trapped coil flux by field comparison with a Pd reference of known susceptibility and sample shape effects by division of an in-plane or out-of-plane filling factor.  

\subsection{PNR}

Measurements were performed at the POLREF beamline at ISIS Neutron and Muon source. Both spin channels were collected by time of flight detection. The 20~mm $\times$ 20~mm sample was attached to a copper puck and thermocouple by thermal paste. Measurements were conducted \textit{in situ} inside a tube furnace under rough vacuum of $\sim 10^{-3}$ provided by a rotary roughing pump throughout the experiment. The sample would be raised to the desired temperature in increasing 50\degc\ increments, held at this temperature for 5 minutes before switching off the furnace and allowing the sample to return to room temperature. The cooling rate was quickened by running helium gas through the back of the puck. After each anneal step, scattering wave-vectors were collected at three angles: 0.5, 1.2 and 2.3$^{\circ}$ for 1, 2 and 8 hours respectively. Data reduction was performed with 10 hours of transmission statistics. 

Slab model fits were modelled using the neutron fitting software Refl1D and fitted using the DREAM algorithm \cite{refl1d}. The model fitted coupled 20 and 350\degc\ slab models which shared identical nuclear SLD distributions but had independent magnetic models. The combined fit achieves a final chi-squared goodness of fit of $\chi^2 = 6.787 \pm 0.099$. The code used is included in the open access repository and builds on that used in the previous study \cite{Brereton2026}.

\subsection{X-PEEM}

All XMCD X-PEEM measurements were conducted at the I06 beamline at the Diamond Light Source. Photons were tuned to the Co L$_3$ edge by XAS measurement and images were collected at room temperature and under a 20~kV bias.

\section{Antoine Vapour Pressure Estimation}

The vapour pressure both of \BS\ and elemental Se was estimated from fitted values from wider literature of the Antoine equation. This equation has the form:
\begin{equation}
    \log_{10}(P) = A - B/(T + C)
\end{equation}
and was used as a reference for deciding the annealing pressure. We note this relation is entirely phenomenological and should only be used as a rough indication in any optimisation process.

\begin{figure}
    \centering
    \includegraphics[width=10cm]{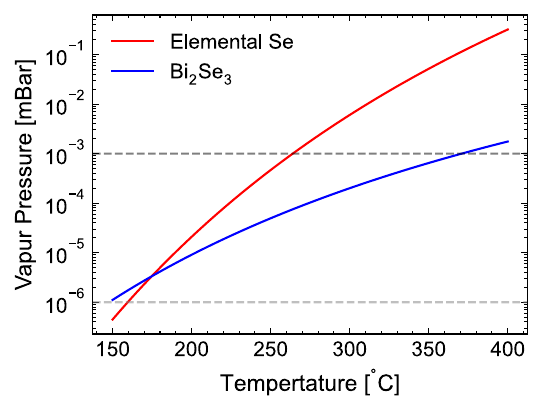}
    \caption{Plotted Antoine fittings for both \BS\ and elemental Se sourced from \cite{Cafaro1984} and \cite{Stull1947}, respectively. $10^{-3}$ and $10^{-6}$ mbar isobars are plotted for reference.}
    \label{fig:Antoine}
\end{figure}

Values used are shown in Table \ref{tab:values} and the resultant relations are plotted in Fig. \ref{fig:Antoine}. We note here that the pressure chosen for the annealing experiment ($\sim 10^{-3}$ mBar) lies above the Se vapour pressure until $\sim 250 $\degc. This is consistent with findings that at the given temperature, the samples can be safely annealed up to $\sim 200$\degc. The process pressure additionally lies above the \BS\ vapour pressure for the whole temperature range used -- consistent with observation that the \BS\ structure is broadly retained. 

\begin{table}[b]
    \centering
    \caption{Antoine values from literature used in equation (1). The plotted relations can be seen in Fig.~\ref{fig:Antoine}.}
    \label{tab:values}
        \begin{tabular}{c| c c c} 
    \hline
    Material &   A [mBar] & B [mBar/K] & C [K]   \\ \hline
        \BS \cite{Cafaro1984}  &  9.325 &  5779 & 270 \\
        Elemental Se \cite{Stull1947} & 6.33714 &  6588.125 & 356.633 \\
    \hline
    \end{tabular}
\end{table}

\section{XRD Analysis}

\begin{figure}
    \centering
    \includegraphics[width=15cm]{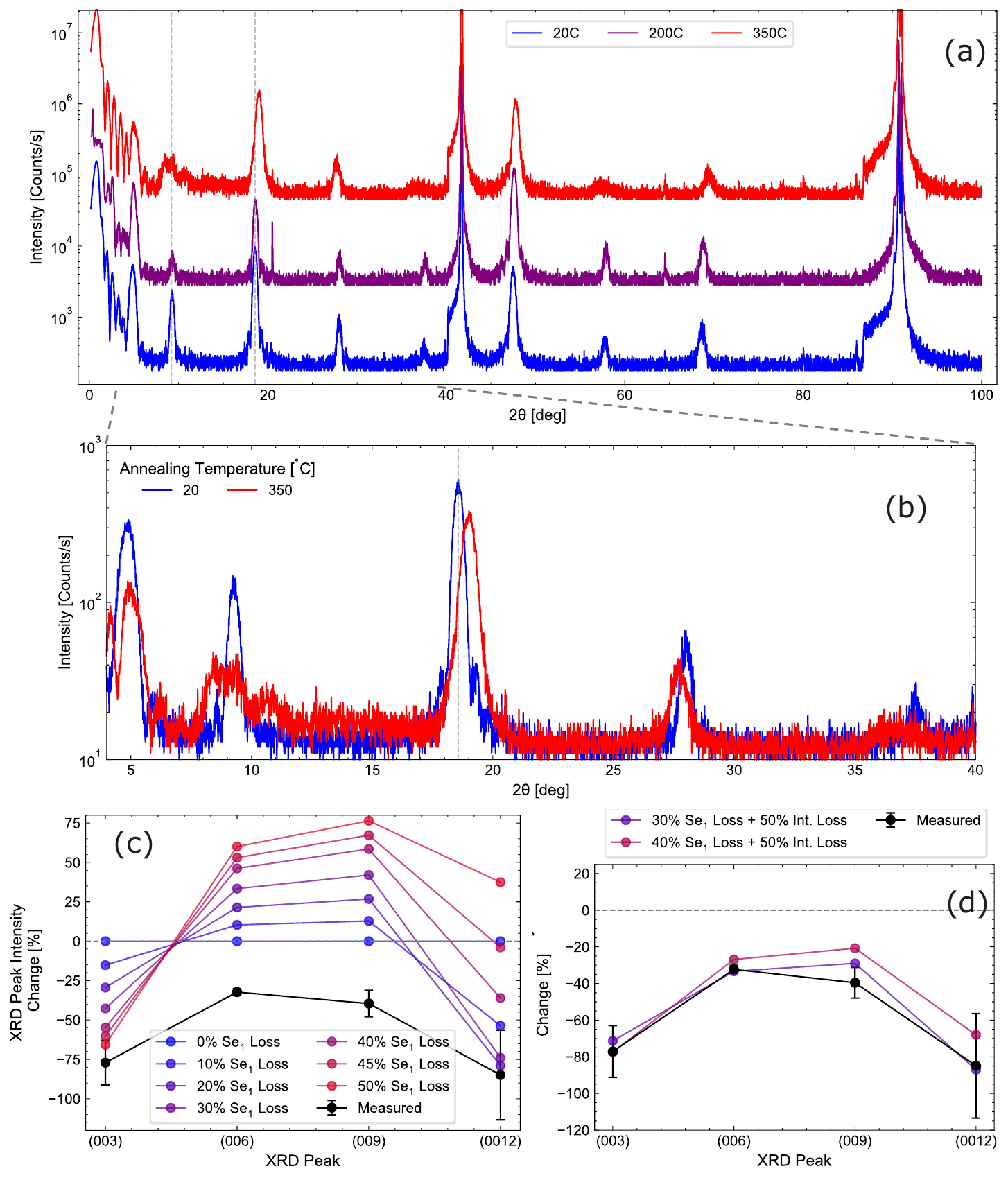}
    \caption{XRD measurements: (a) Patterns for samples as grown, after a 200\degc\ anneal, and after 350\degc\ anneal; (b) superposed low angle XRD data for identical sample pre- and post-anneal overlaid and normalised to substrate peak; (c) the fractional decrease in peak intensity from annealing to 350\degc\ compared to those predicted by structure factor calculations from a 10\% loss to the Se$_1$ site; (d) a rough match to the measured data obtained by combining 30 - 40\% site specific loss from (c) with an overall 50\% loss of peak intensity.}
    \label{fig:XRD}
\end{figure}

All x-ray measurements were taken using a Cu K-$\alpha$ source with $\lambda = 1.54$~\AA\ on a Bruker D8 Discovery X-ray diffractometer. The presence of only $(00l)$ XRD peaks confirms that \BS\ grows predominantly along the $c$-axis. These peaks were seen to have changed substantially after annealing as is shown in Fig.~\ref{fig:XRD}(a-b). Noticeably, all peaks were seen to reduce in intensity however the percentage change disproportionately effected the $(003)$ and $(0\,0\,12)$ peaks.

\begin{figure}
    \centering
    \includegraphics[width=12cm]{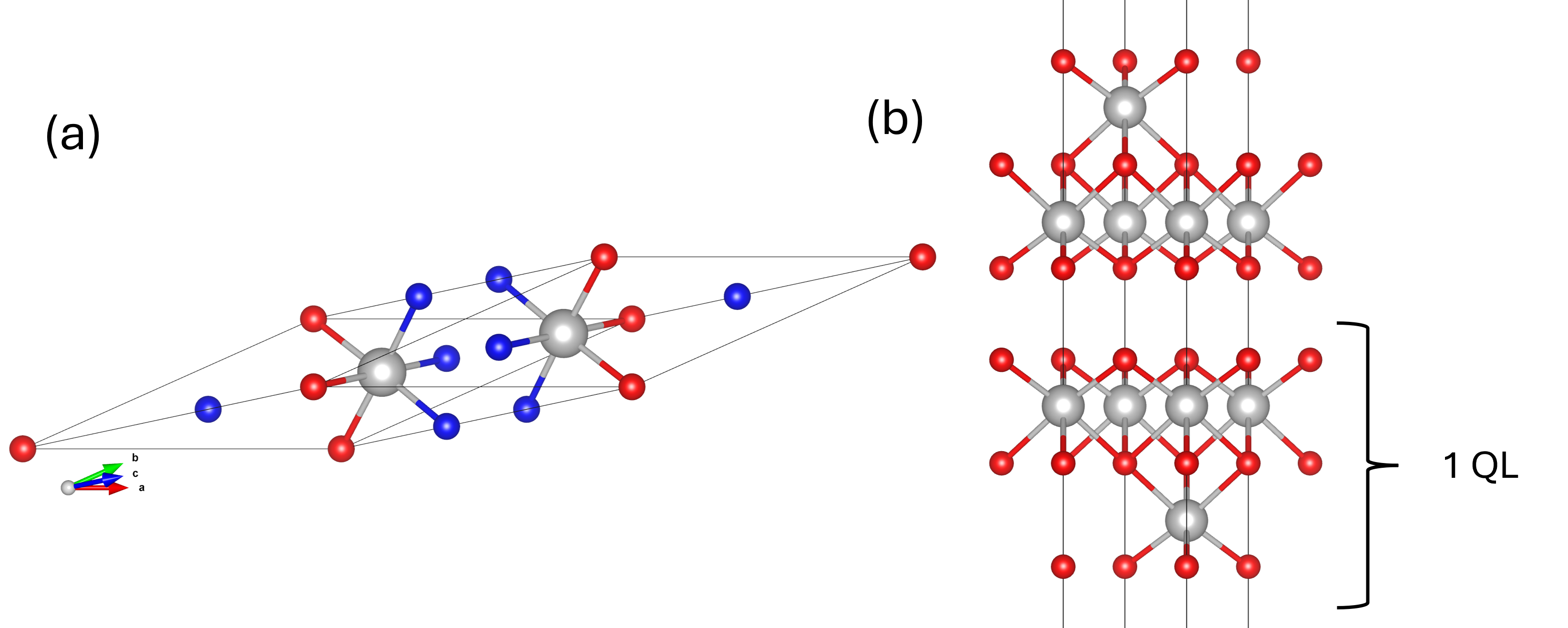}
    \caption{\BS\ represented as both a primitive $R3m$ rhombohedral, unit cell (a) and in the conventional standard hexagonal representation (b) where the quintuple layering can be seen. Coloured sites denote Se. In (a) the Se$_1$ sites used in the structure factor calculations are marked as blue. Figures were rendered in VESTA \cite{Momma2011}.}
    \label{fig:vesta}
\end{figure}

This qualitatively agrees with a simple structure factor calculation performed using the VESTA software package \cite{Momma2011} which shows the same peaks reducing due to a loss of Se occupancy on the Se$_1$ site of the $R3m$ unit cell as shown in Fig.~\ref{fig:vesta}(a). Notably a modelled uniform loss of Se throughout the unit cell would correspond to a relative increase of the $(0\,0\,12)$ peak over the $(006)$. It hence appears the Se loss is concentrated on the Se$_1$ site which is consistent with first principles calculations that suggest that vacancies at the Se$_1$ sites are energetically favourable due to weak bonding in the van der Waals gap \cite{Xue2013}. Absolute values for the peak intensity reduction differ substantially however. Considering all peaks experience a substantial reduction this can be explained by an additional reduction in \BS\ crystallinity and applying a 50\% reduction to the modelled values for 30 - 40\% Se$_1$ loss offers fairly close agreement to the observed values as shown in Fig.~\ref{fig:XRD}(d). 

We see also in Fig.~\ref{fig:XRD} that the peak positions shift, which we attribute to changes in the size of the unit cell due to Se loss. The $(006)$ peak is the only peak to shift rightwards. Given its particular sensitivity to the Bi-Bi spacing, we attribute this to the Bi spacing decreasing whilst the Se spacings relax. We also observed a loss of all Laue oscillations on the $(006)$ peak after annealing to 350\degc\ as seen in Fig.~\ref{fig:XRD}(b), which reinforces the loss of epitaxy in our films.  

\bibliographystyle{iopart-num}
\bibliography{library}